\documentclass[trackchanges,twocolumn]{aastex702}

\newcommand{\civ}{\ion{C}{4}}
\newcommand{\heii}{\ion{He}{2}}
\newcommand{\niii}{\ion{N}{3}}

\newcommand{\hb}{H\ensuremath{\beta}}

\begin{document}
\title{Identification of the \civ~$\lambda\lambda$5801, 5812 Emission in the Off-Nuclear TDE 2025abcr}
\author[orcid=0000-0003-3472-4392,sname='Sun']{Jingbo Sun}
\affiliation{Shanghai Astronomical Observatory, Chinese Academy of Sciences, 80 Nandan Road, Shanghai 200030, People's Republic of China}
\affiliation{University of Chinese Academy of Sciences, 19A Yuquan Road, 100049, Beijing, People's Republic of China}
\affiliation{Leiden Observatory, Leiden University, Postbus 9513, 2300 RA, Leiden, The Netherlands}
\email[show]{sunjingbo@shao.ac.cn}  

\author[orcid=0000-0001-8416-7059, sname='Guo']{Hengxiao Guo} 
\affiliation{Shanghai Astronomical Observatory, Chinese Academy of Sciences, 80 Nandan Road, Shanghai 200030, People's Republic of China}
\email[show]{hengxiaoguo@gmail.com}

\author[orcid=0000-0002-3859-8074]{Sjoert van Velzen} 
\affiliation{Leiden Observatory, Leiden University, Postbus 9513, 2300 RA, Leiden, The Netherlands}
\email{sjoert@strw.leidenuniv.nl}

\author[orcid=0000-0001-6747-8509]{Yuhan Yao} 
\affiliation{Kavli Institute for Astronomy and Astrophysics, Peking University, Beijing, 100871, People's Republic of China}
\email{yyao@pku.edu.cn}

\author[orcid=0000-0003-4440-259X]{Mar Mezcua}
\affiliation{Institute of Space Sciences (ICE, CSIC), Campus UAB, Carrer de Magrans, 08193 Barcelona, Spain}
\affiliation{Institut d'Estudis Espacials de Catalunya (IEEC), Edifici RDIT, Campus UPC, 08860 Castelldefels (Barcelona), Spain}
\email{mezcua@ice.csic.es}

\author[orcid=0000-0001-9456-8358]{Stefan Geier}
\affiliation{GRANTECAN, Cuesta de San Jos\'e s/n, 38712 Bre\~na Baja, La Palma, Spain}
\affiliation{Instituto de Astrof\'isica de Canarias, V\'ia L\'actea, 38205 La Laguna, Tenerife, Spain}
\email{stefan.geier@gtc.iac.es}

\author[0000-0002-9331-4388]{Alok C. Gupta}
\affiliation{Aryabhatta Research Institute of Observational Sciences (ARIES), Manora Peak, Nainital-263001, India} 
\affiliation{Xinjiang Astronomical Observatory, Chinese Academy of Sciences, 150 Science-1 Street, Urumqi 830011, People's Republic of China}
\email{acgupta30@gmail.com}

\author[0009-0007-3214-602X]{Karan Dogra}
\affiliation{Aryabhatta Research Institute of Observational Sciences (ARIES), Manora Peak, Nainital-263001, India} 
\email{karandogra987@gmail.com}

\author[0000-0002-8186-4753]{Philip G. Edwards} 
\affiliation{CSIRO Space and Astronomy, PO Box 76, Epping, NSW, 1710, Australia}
\email{philip.edwards@csiro.au}

\author[0000-0002-4455-6946]{Wenwen Zuo}
\affiliation{Shanghai Astronomical Observatory, Chinese Academy of Sciences, 80 Nandan Road, Shanghai 200030, People's Republic of China}
\email{wenwenzuo@shao.ac.cn}

\author[0009-0002-7625-2653]{Yu Pan}
\affiliation{South-Western Institute for Astronomy Research, Yunnan Key Laboratory of Survey Science, Yunnan University, Kunming, Yunnan 650500, People's Republic of China}
\email{yupan_astro@outlook.com}

\author[0000-0002-4455-6946]{Minfeng Gu}
\affiliation{Shanghai Astronomical Observatory, Chinese Academy of Sciences, 80 Nandan Road, Shanghai 200030, People's Republic of China}
\email{gmf@shao.ac.cn}

\begin{abstract} 
TDE~2025abcr is a tidal disruption event (TDE) located approximately $9$\,kpc in projection from its host-galaxy nucleus. Our spectroscopic follow-up with the Very Large Telescope (VLT) reveals \civ~$\lambda\lambda5801, 5812$ emission, representing the first identification of this \civ\, doublet in a TDE. These transitions are uncommon in active galactic nuclei (AGN) spectra but have been observed in supernova flash spectra and Wolf--Rayet (WR) stars. We propose that the doublet arises from the combined effects of continuum fluorescence and photon trapping in the $312$\,\AA\ transition, analogous to the mechanism operating in nitrogen-sequence Wolf--Rayet (WN) winds. This mechanism requires intense extreme-ultraviolet (EUV) illumination and a high optical depth, providing a distinctive diagnostic of the physical conditions in optically thick TDE outflows. These requirements, together with the relatively narrow widths of the broad emission lines, the presence of Bowen fluorescence features, and the small inferred blackbody radius, favor a compact and dense envelope in TDE~2025abcr. The optical \civ\ doublet may have been overlooked in previous observational and theoretical studies of TDEs, motivating more detailed radiative-transfer calculations for optically thick outflows. With future TDE samples from Rubin/LSST and WFST, a larger \civ\, TDE sample will test if this population systematically favors a compact photosphere. 
\end{abstract}

\section{Introduction}
\label{sec:intro}
A tidal disruption event (TDE) occurs when a star passes sufficiently close to a massive black hole (MBH) to be torn apart by tidal forces, producing a luminous flare as the bound debris returns and accretes onto the MBH \citep{Rees1988}. Modern wide-field time-domain surveys have increased the number of known optical TDEs to more than 100, enabling systematic studies of their population properties \citep{vanVelzen2021,Hammerstein2023,Zhang2026}. Complementing studies of sustained accretion in active galactic nuclei (AGN), TDEs reveal otherwise quiescent MBHs in inactive galaxies \citep{Gezari2021,Wevers2017,Yao2023} and provide laboratories for testing accretion theory across rapidly changing accretion rates \citep{Strubbe2009,Wevers2021,Sun2025,Yao2026}.

Optically selected TDEs typically exhibit diverse emission-line properties and spectral evolution. Based on the presence of transient hydrogen lines or helium lines, TDE spectra are commonly classified as TDE-H, TDE-He, TDE-H+He, or TDE-featureless \citep{vanVelzen2021,Hammerstein2023}. A subset of TDE-H+He spectra also exhibits \niii~$\lambda4640$ and O~III~$\lambda3760$ emission associated with Bowen fluorescence, revealing illumination by an extreme-ultraviolet (EUV) source \citep{Blagorodnova2019,Leloudas2019}. These Bowen TDEs tend to have smaller blackbody radii and higher temperatures \citep{vanVelzen2021}, together with narrower emission lines than hydrogen-dominated events \citep{Charalampopoulos2022}, suggesting a connection between spectral diversity and the structure of the emitting gas. More generally, TDE line widths range from several thousand to $\sim10^{4}$\,km\,s$^{-1}$ and often decrease as the flare fades, a behavior that can be explained by electron-scattering broadening in an evolving reprocessing layer \citep{Roth2018}. Line strengths and ratios further probe the excitation conditions, radiative-transfer effects, and elemental abundances, although these contributions can be difficult to disentangle \citep{Roth2016,Thomsen2026}. 


An emerging population of off-nuclear TDEs extends these studies to MBHs outside galactic centers. Examples include X-ray TDEs 3XMM~J2150 \citep{Lin2018} and EP240222a \citep{Jin2025}, and optical TDEs TDE~2024tvd \citep{Yao2025}, TDE~2023mfm \citep{Sfaradi2026}, and TDE~2025abcr \citep{Stein2026}. Among optically selected TDEs, TDE~2024tvd and TDE~2023mfm have projected nuclear offsets of $0.8$ and $1.1$,kpc, respectively, placing them within the bright bulge regions of their hosts. By comparison, TDE~2025abcr lies at a projected distance of $9.3$\,kpc from the nucleus of its host galaxy at $z\simeq0.05$. The substantially lower host-galaxy contamination enables detailed measurements of emission-line evolution at later post-peak epochs. TDE~2025abcr ($01^{\mathrm{h}}46^{\mathrm{m}}55.40^{\mathrm{s}}$, $-15^\circ22^\prime15.7^{\prime\prime}$) was first detected by the Zwicky Transient Facility (ZTF) on 2025 October 13 and subsequently confirmed as a H+He Bowen TDE through spectroscopic follow-up.

Beyond its off-nuclear location, TDE~2025abcr exhibits relatively narrow broad emission lines with widths of $\sim2500$\,km\,s$^{-1}$ \citep{Stein2026}, potentially indicating a lower optical depth in the electron scattering region \citep{Roth2018}. The narrow line width bears similarity with the EP240222a, a candidate intermediate-mass black hole (IMBH) TDE that exhibits line widths of $\sim1000$\,km\,s$^{-1}$ \citep{Jin2025}. Together, the large projected nuclear offset and the narrow emission line width make TDE~2025abcr a valuable target for investigating a possible IMBH origin. Its peak luminosity yields a BH mass of order of $10^{6}\,M_{\odot}$ \citep{Stein2026}, while this estimate carries substantial uncertainty \citep{Mummery2024}, motivating further investigation of the TDE emission structure to understand the underlying physical processes.


In this Letter, we present photometric and spectroscopic monitoring of TDE~2025abcr and report the identification of the optical \civ~$\lambda\lambda5801, 5812$ doublet (also referred to as the \civ~$\lambda5805$ emission) in a TDE. We investigate its excitation and implications for the line-emitting gas, and report a complementary search for intranight optical variability that yields no significant detection. Section~\ref{sec:observations} describes the observations and data reduction. Section~\ref{sec:results} presents the light-curve evolution, spectral measurements, and intranight-variability constraints. Section~\ref{sec:diss} discusses the origin of the \civ\, emission and the implications for the TDE envelope. We adopt a flat $\Lambda$CDM cosmology with $H_0=70$\,km\,s$^{-1}$\,Mpc$^{-1}$, $\Omega_{\rm M}=0.3$, and $\Omega_{\Lambda}=0.7$.

\begin{figure}[t]
  \centering
  \includegraphics[width=\columnwidth]{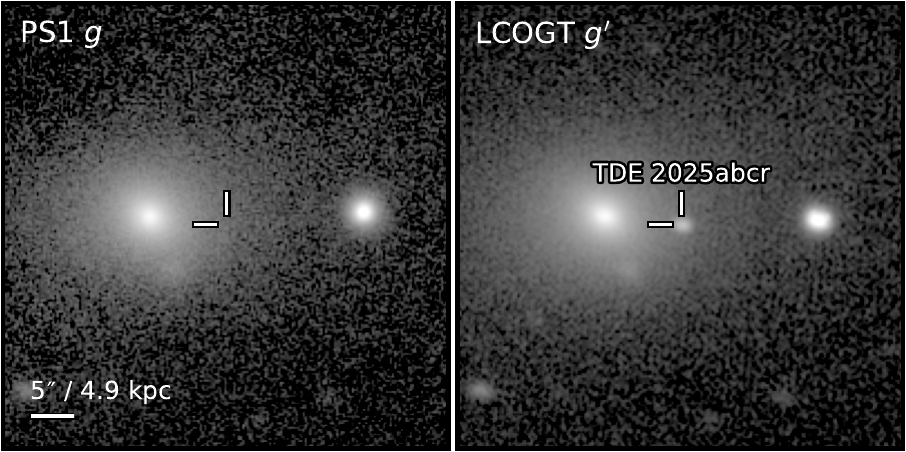}
  \caption{The archival PS1 and the LCOGT $g$-band images of TDE~2025abcr.}
  \label{fig:image}
\end{figure}
 
\section{Observations and Data Reduction}
\label{sec:observations}

\subsection{Optical Light Curves}
\label{sec:lc}
We obtained the ZTF $g$- and $r$-band forced photometry at the position of TDE~2025abcr through the ZTF forced-photometry service \citep{Masci2023}, which performs point-spread-function (PSF) fitting at fixed sky coordinates on difference images generated by the ZTF Science Data System \citep{Masci2019}. Nondetections are reported as $3\sigma$ upper limits.

\begin{figure*}[ht!]
    \centering
    \includegraphics[width=\linewidth]{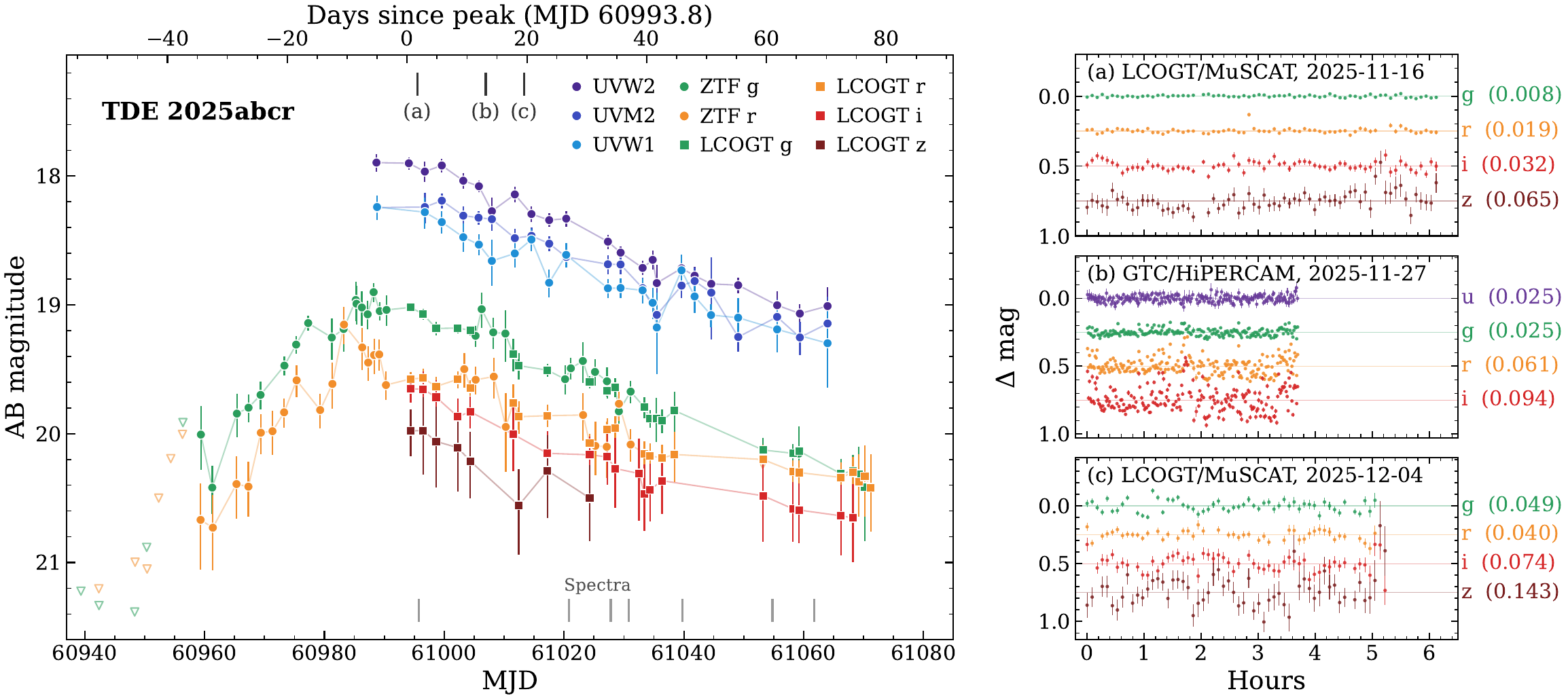}
    \caption{UV and optical photometric evolution of TDE~2025abcr. Left: multi-band light curves from ZTF, LCO, and \textit{Swift}/UVOT, with filters and instruments distinguished by color and symbol, respectively. Downward-pointing triangles indicate $3\sigma$ upper limits. Tick marks at the top indicate the three high-cadence photometric monitoring epochs, while those at the bottom mark the spectroscopic epochs. Right: differential light curves from the individual monitoring sequences, vertically offset for clarity. The variability limit (in magnitudes) is indicated alongside each band label.}
    \label{fig:lc}
\end{figure*}

We monitored TDE~2025abcr with the Las Cumbres Observatory Global Telescope network (LCOGT), using MuSCAT3 of the Faulkes Telescope North (FTN) at Haleakala Observatory and MuSCAT4 of the Faulkes Telescope South (FTS) at Siding Spring Observatory. Simultaneous 300\,s exposures were obtained in the $g'$, $r'$, $i'$, and $z_s$ bands, hereafter denoted $griz$. We use the images reduced by the BANZAI pipeline \citep{McCully2018}, create difference images using \texttt{HOTPANTS} \citep{Becker2015}, and perform forced PSF photometry at the transient position using \texttt{photutils}. The absolute transient flux in the LCOGT reference image is measured by subtracting the PS1 image and then added to the differential light curves to recover the total transient flux. Flux uncertainties include contributions from PSF photometry and the reference-flux measurement. The image is shown in Figure~\ref{fig:image}.

\subsection{Intranight Monitoring}
\label{sec:intranight}
To search for hour-scale variability and perform continuum reverberation mapping to measure the BH mass \citep{Fausnaugh2016,Guo2026}, a 5\,hr monitoring sequence was obtained with HiPERCAM on the Gran Telescopio Canarias (GTC) on 2025 November~27 through Director's Discretionary Time (program DDT GTC2025-289, P.I. Mezcua). HiPERCAM provides simultaneous five-band imaging through filters similar to the SDSS $ugriz$ bands \citep{Dhillon2021}, with a native pixel scale of $0.081\,\mathrm{arcsec\,pixel^{-1}}$ and a full-frame field of view of approximately $2.8\times1.4$\,arcmin. The observations used $2\times2$ CCD binning, yielding a pixel scale of $0.162\,\mathrm{arcsec\,pixel^{-1}}$, with integration times of 32\,s in $u$ and 4\,s in $griz$. The data were reduced using the HiPERCAM pipeline. Aperture photometry was performed with a radius of $1.2$ times the measured PSF full width at half maximum (FWHM), and the sky background was estimated from an annulus spanning $2.5$--$3.0$ times the FWHM, masking regions contaminated by the host galaxy. Differential light curves were constructed relative to a comparison star. The seeing remained approximately $1$--$2$\,arcsec for most of the sequence but deteriorated during the final 1.5\,hr. These exposures were excluded because of poor image quality and unreliable separation of the transient and host-galaxy flux. Aperture placement for the transient and comparison stars was visually verified in all retained exposures.

Additional continuous monitoring was obtained with LCOGT using consecutive 300\,s exposures, spanning approximately 6\,hr on 2025 November~16 and 5\,hr on 2025 December~4. These observations were reduced following the procedure described above.

\begin{figure*}
    \centering
    \includegraphics[width=\linewidth]{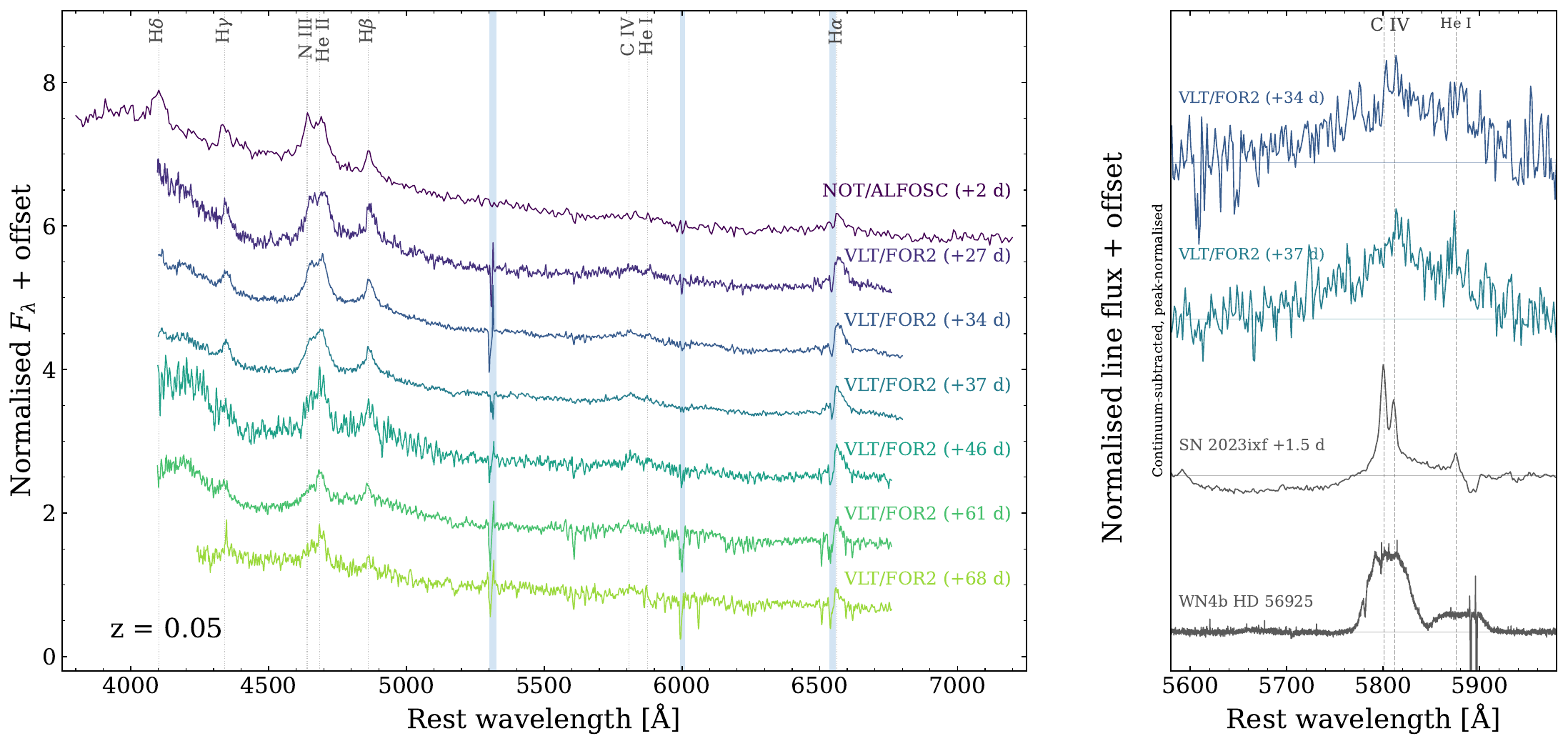}
    \caption{
    Spectral evolution of TDE~2025abcr. Left: the NOT and VLT spectra, vertically offset for clarity. Vertical markers identify the principal emission features, while shaded intervals indicate the regions contaminated by the telluric absorption. Right: enlarged views of the \civ/\ion{He}{1} region for two representative VLT spectra. An early-time spectrum of SN~2023ixf and a WN-star spectrum are shown for comparison. All the phases are the relative time to the optical peak in the rest frame.
    }
    \label{fig:spectra}
\end{figure*}

\subsection{Swift}
\label{sec:swift}
TDE~2025abcr was observed with the X-Ray Telescope (XRT) and Ultraviolet/Optical Telescope (UVOT) onboard the \textit{Neil Gehrels Swift Observatory}. We obtained XRT light curves from the UK Swift Science Data Centre online product generator \citep{Evans2009}, selecting 0.3 -- 10~keV and a minimum of 10 counts per bin. A faint X-ray source is detected near the transient, consistent with the reports of \citet{Stein2026} and \citet{Patra2026}, with a luminosity of approximately $10^{42}$\,erg\,s$^{-1}$ and no evidence for intrinsic absorption. 
Details of the X-ray analysis will be presented by Y. Zhang et al (in prep). We extracted UVOT photometry in the UVW2, UVM2, and UVW1 filters using \texttt{uvotsource}. Source counts were measured in a circular aperture of radius $5$~arcsec, with the background estimated from a nearby source-free region. The UV images show a point source at the transient position without obvious extended host emission. We excluded the $u$-band measurements because of their low signal-to-noise ratio (SNR) and host-galaxy contamination. The UV photometry was corrected for Galactic extinction using $E(B-V)=0.013$~mag \citep{Schlafly2011}.

\subsection{Spectra}
\label{sec:spec}
Six epochs of optical spectroscopy were obtained with FORS2 on the Very Large Telescope (VLT) through Director's Discretionary Time (program 116.2APM, PI: Sun). The observations used the 600V+94 grism and a $1.3$\,arcsec slit, providing nominal observed-frame wavelength coverage of 443--737\,nm and a resolving power of $R\simeq900$. The spectra were reduced using the ESO Data Processing System (EDPS) and the FORS2 long-slit spectroscopy pipeline.


We also obtained one spectrum on 2025 November~17 with the Alhambra Faint Object Spectrograph and Camera (ALFOSC) mounted on the Nordic Optical Telescope (NOT), with two 2800~s exposures using grism~4 and a $1.3$~arcsec slit, yielding a spectral resolving power of $R\sim300$. The data were reduced using the standard pipeline. All the extracted spectra are scaled to contemporaneous photometry, and corrected for Galactic extinction following \citet{Fitzpatrick1999}.

\section{Results}
\label{sec:results}

\subsection{Light curves}
\label{sec:lightcurves}
Figure~\ref{fig:lc} presents the UV and optical light curves of TDE~2025abcr, together with the three sequences of high-cadence photometry. We adopt $t_{\rm peak}=\mathrm{MJD}\,60993.8$ (2025 November~14), determined from g-band light curve fit. The spectral energy distributions (SEDs) is fitted with a single-temperature blackbody, obtaining $T_{\rm BB}\simeq2\times10^4$\,K with no significant evolution over the monitored interval. This temperature is slightly lower than the $3\times 10^{4}$ K fitted from UV and blue optical bands \citep{Stein2026}. This discrepancy may arise because the $i$- and $z$-band flux densities systematically exceed the blackbody predictions constrained by the UV and blue optical bands. This excess suggests an additional emission component, such as free–free emission \citep{Reynolds2026,Wu2026,Patra2026}, or a departure from a single-temperature continuum, as expected from a multi-temperature photosphere \citep{Guo2025}.

\begin{figure*}[ht!]
    \centering
    \includegraphics[width=\linewidth]{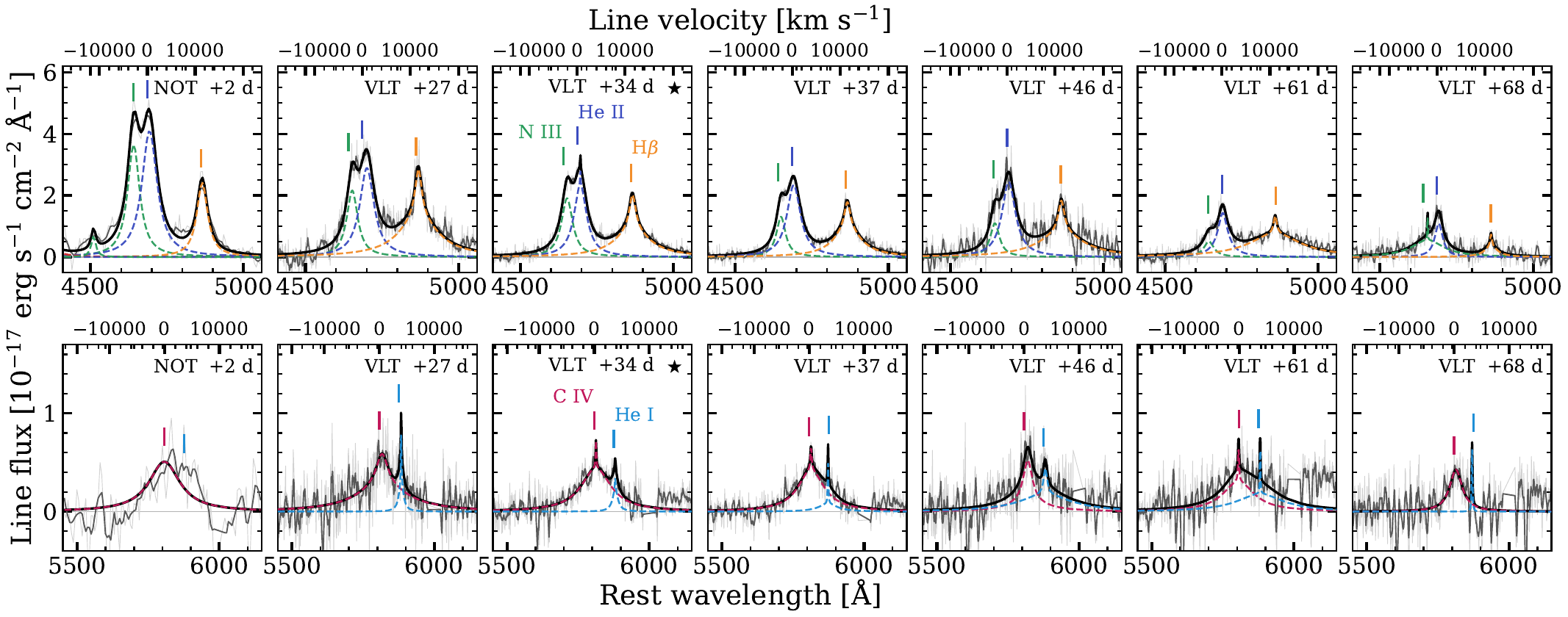}
    \caption{
    Emission-line fits to the continuum-subtracted spectra of TDE~2025abcr. The upper panels show the \ion{N}{3}, \ion{He}{2}, and H$\beta$ region, while the lower panels show the \civ\, and \ion{He}{1} region. Black curves show the spectra smoothed over five pixels, with the original spectra shown in gray. Red curves represent the total fitted emission-line models, and colored dashed curves show the individual components. Vertical markers indicate the reference wavelengths. Velocities are measured relative to \ion{He}{2} $\lambda4686$ in the upper panels and \civ~$\lambda5805$ in the lower panels. Star symbols mark the epochs with the highest-quality spectra, where the \civ\, and \ion{He}{1} lines are well separated. 
    }
    \label{fig:linefits}
\end{figure*}

\begin{figure}[h!]
    \centering
    \includegraphics[width=\linewidth]{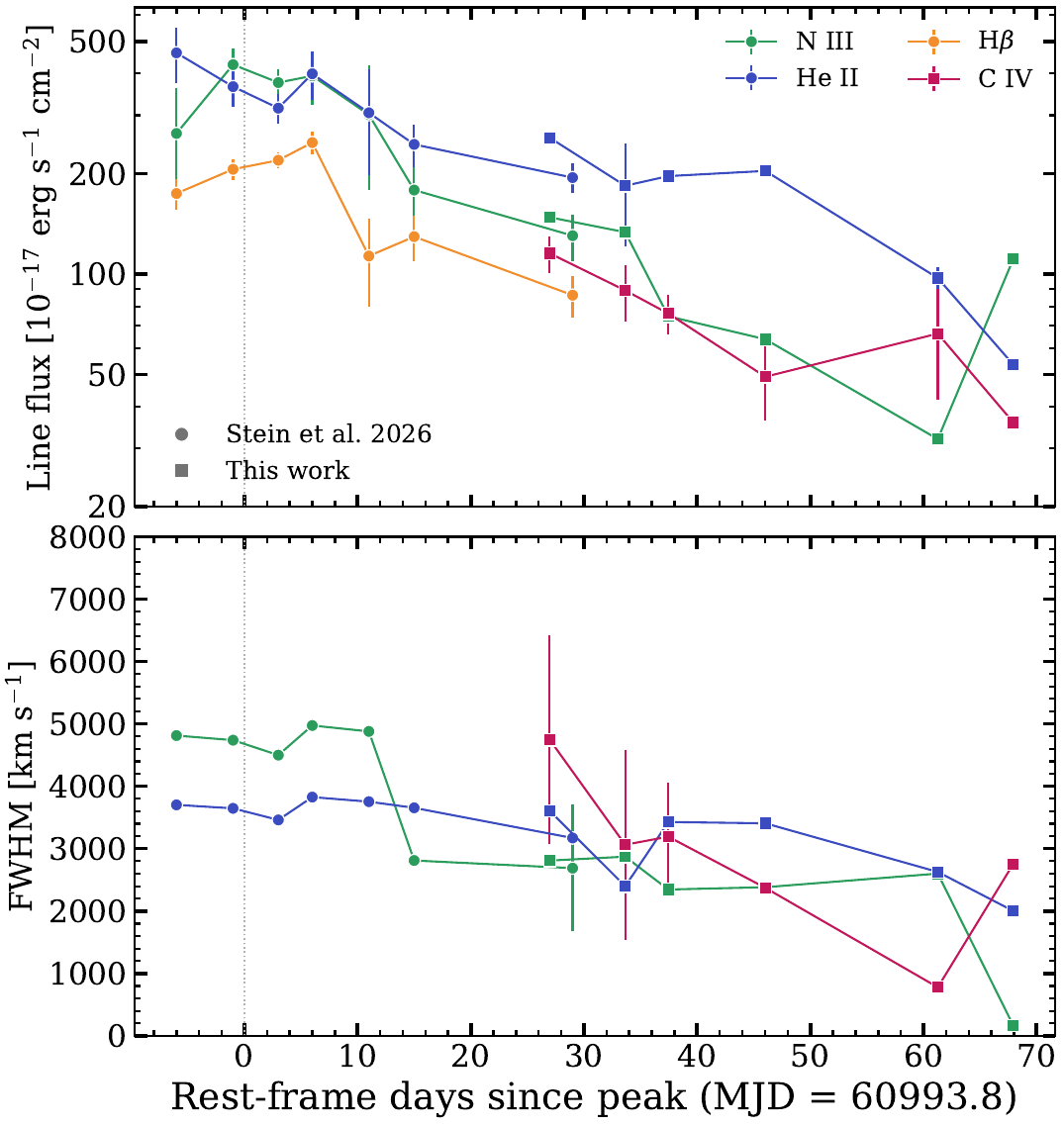}
    \caption{
    Evolution of the integrated emission-line fluxes (upper panel) and FWHM values (lower panel). The vertical dotted line marks the adopted peak epoch. Early-time \civ\, and late-time H$\beta$ measurements are omitted because the SNR are insufficient for reliable continuum subtraction and emission-line fitting. Circles denote measurements from fits to the spectra presented by \citet{Stein2026}, while squares denote those from the spectra obtained in this work. 
    }
    \label{fig:line_evolution}
\end{figure}

High-cadence observations on 2025 November~16, November~27, and December~4 revealed no statistically significant intranight variability. The most stringent upper limits on hour-scale variability were $0.008$\,mag in $g$ and $0.025$\,mag in $u$, with weaker constraints in the remaining bands (Figure~\ref{fig:lc}). The absence of rapid variability is consistent with emission from an extended, reprocessing-dominated optical photosphere with an inferred radius of $\sim10^{14}$\,cm, as commonly invoked for nuclear TDEs with $10^{5-7}\ M_{\odot}$  \citep{Holoien2019}, although it does not independently constrain the BH mass.

\subsection{Spectra}
\label{sec:spectra}
Figure~\ref{fig:spectra} presents the spectral time-sequence of TDE~2025abcr. In addition to the established TDE features of Balmer emission, \heii~$\lambda4686$, \niii~$\lambda4640$, and \ion{He}{1} features, we identify a new emission feature near $5805$\,\AA\, in both our VLT spectra and those presented by \citet{Stein2026}. Its wavelength is consistent with the blended \civ~$\lambda\lambda5801, 5812$ doublet, and its peak is distinct from the neighboring \ion{He}{1}~$\lambda5876$ emission. Alternative identifications as \ion{He}{1}~$\lambda5876$ or [\ion{N}{2}]~$\lambda5754$, as discussed in \citet{Leloudas2019}, would require a blueshift of $\sim3600$\ km\ s$^{-1}$ or a redshift of $\sim2600$\ km\ s$^{-1}$, respectively. By comparison, the \heii\ and \niii\ centroids in TDE~2025abcr show redshifts of $\sim10^{3}$\ km\ s$^{-1}$, providing no evidence for comparable velocity offsets. The wavelength agreement and separation from the neighboring \ion{He}{1} peak therefore favor the \civ\ identification. The excitation of this feature is discussed in Section~\ref{sec:diss}.

To quantify the emission features, we perform the emission line fit for the \niii/\heii\ regions and \civ/\ion{He}{1} regions separately (Figure~\ref{fig:linefits}). In each region, the local continuum is modeled with a blackbody function, with a power-law prescription yielding similar results. Following continuum subtraction, the emission lines were fitted simultaneously, with each line modeled as the sum of two Lorentzian components sharing a common centroid to describe the peaked profiles and broad wings characteristic of TDE emission lines \citep{Charalampopoulos2022}. The reported integrated fluxes and FWHM values were measured from the summed profile of each line. The same procedure was applied to the spectra presented by \citet{Stein2026}.

Figure~\ref{fig:line_evolution} shows the evolution of the fitted line fluxes and widths. As the transient fades, the \civ\, doublet and other prominent emission lines generally weaken and narrow. The \heii\ and \niii\ FWHM values decrease from approximately $4000$--$5000$\,km\,s$^{-1}$ near maximum to $\sim2000$\,km\,s$^{-1}$ in the final spectrum. The comparable widths of \civ\, and the other lines suggest that they experience similar scattering conditions. The overall line narrowing with decreasing luminosity is consistent with the electron scattering contribution to the line profiles. The limited SNR and uncertainty in the continuum shape cannot support a more detailed assessment of the line profile evolution, particularly for \hb\ in the VLT spectra and \civ\, at early epochs.

\begin{figure}[]
    \centering
    \includegraphics[width=\linewidth]{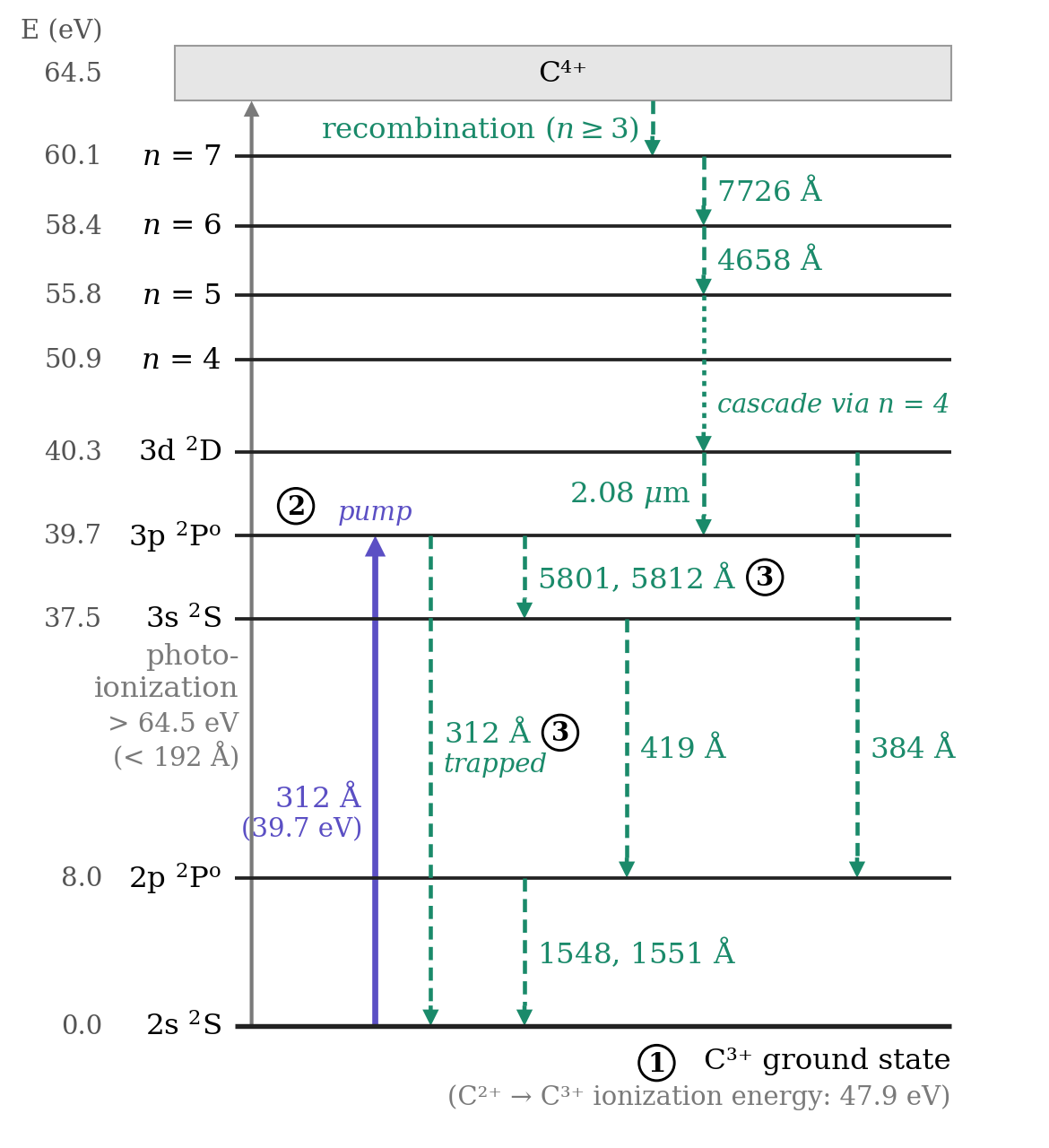}
    \caption{
    Schematic of the excitation and radiative decay pathways of C$^{3+}$. The gray arrow denotes photoionization to C$^{4+}$ in the recombination scenario, the blue arrow denotes continuum pumping near $312$\,\AA, and the green dashed arrows indicate radiative decay pathways. The numbered labels identify three stages relevant to fluorescence-powered optical \civ\, emission: (1) formation of C$^{3+}$ through photoionization; (2) excitation from the $2s$ ground state to the $3p$ levels by absorption of EUV continuum photons; and (3) trapping of photons in the $312$\,\AA\ transition, which enhances conversion into the \civ~$\lambda\lambda5801, 5812$ doublet through the alternative $3p\rightarrow3s$ decay channel. Subsequent decay through the $2p$ levels produces the ultraviolet \civ~$\lambda\lambda1548,1551$ doublet. Selected recombination-cascade pathways are also shown for comparison.
    }
    \label{fig:channel}
\end{figure}

\section{Discussion and Conclusion}
\label{sec:diss}

\subsection{\civ\ Emission Mechanism}
\label{subsec:mechanism}
To our knowledge, this is the first reported identification of the optical \civ~$\lambda\lambda5801, 5812$ doublet in a TDE. These transitions are also uncommon in AGN spectra but have previously been observed in supernova flash spectra \citep{Jacobson-Galan2023,Bostroem2023} and Wolf--Rayet (WR) stars \citep{Crowther2007}. A pure recombination interpretation of the observed \civ/\heii\, line ratio would require substantial carbon enrichment, comparable to that in carbon-sequence WR (WC) stars. Such progenitors are expected to be very rare in the quiescent early-type host galaxy of TDE 2025abcr \citep{Stein2026}, making this scenario less plausible.

For the TDE \civ~$\lambda5805$ emission, we propose a mechanism established in the winds of nitrogen-sequence Wolf--Rayet (WN) stars, in which continuum fluorescence and photon trapping jointly promote optical \civ\, emission \citep{Hillier1988,Hillier2011}. Figure~\ref{fig:channel} illustrates this mechanism alongside the recombination pathways. The main stages are:
\begin{enumerate}
    \item Photoionization of C$^{2+}$ by photons with energies above $47.9$\,eV establishes a population of C$^{3+}$ ions.
    \item Absorption of EUV continuum photons near $312$\,\AA\ excites C$^{3+}$ from the $2s$ ground state to the $3p$ levels.
    \item The excited ions preferentially decay directly to the ground state, emitting $312$\,\AA\ photons. A large optical depth in this transition traps these photons through repeated absorption and re-emission, providing repeated opportunities for decay through the alternative $3p\rightarrow3s$ channel and thereby enhancing the optical \civ~$\lambda\lambda5801, 5812$ emission.
\end{enumerate}
The subsequent $3s\rightarrow2p\rightarrow2s$ cascade produces emission near $419$\,\AA\ and the observable UV \civ~$\lambda\lambda1548,1551$ doublet, which is commenly seen in TDE UV spectra \citep[e.g.,][]{Cenko2016,Hammerstein2026}. This fluorescence pathway therefore converts incident EUV radiation into emission at longer wavelengths. Other observed properties of TDE~2025abcr provide complementary constraints, as the blackbody fits suggest a relatively compact continuum photosphere, while the relatively narrow emission lines indicate a modest electron-scattering optical depth above the line-forming region. Together, these properties are consistent with a compact and dense envelope in TDE~2025abcr. A larger sample of \civ-emitting TDEs, combined with quantitative radiative-transfer calculations, is needed to determine whether these \civ\ TDEs preferentially exhibit similar properties observed in TDE~2025abcr.

\begin{figure}[t!]
    \centering
    \includegraphics[width=\linewidth]{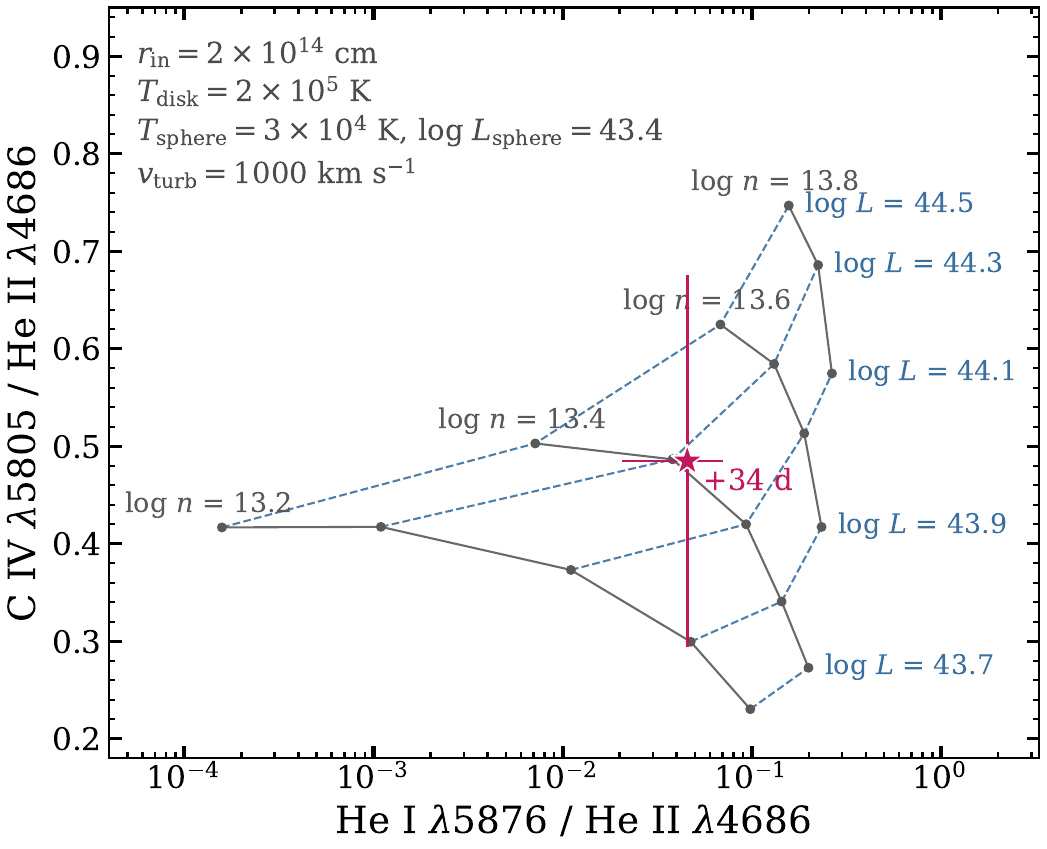}
    \caption{Exploratory \texttt{CLOUDY} calculations of the observed emission-line ratios. The limitations of this toy model are discussed in Section~\ref{subsec:mechanism}. Within the adopted framework, the \civ~$\lambda5805$ emission favors dense gas illuminated by strong EUV radiation. 
    }
    \label{fig:cloudy}
\end{figure}

Motivated by these excitation mechanisms, an exploratory photoionization grid was calculated using \texttt{CLOUDY}  \citep{Chatzikos2023} to examine the conditions that can enhance optical \civ\, emission. We note that \texttt{CLOUDY} cannot simulate the optically thick TDE outflows, while the result shown in Figure~\ref{fig:cloudy} aims to explore qualitative trends for the EUV radiation and the density, rather than to infer the physical parameters or reconstruct the gas structure of TDE~2025abcr. The models adopted two ionization sources: (i) an inner disk with a temperature of $2\times10^{5}$\,K, with luminosity varying from $10^{43.7 - 44.5}\rm\ erg\ s^{-1}$, and (ii) a hot photosphere with the blackbody temperature of $3\times 10^{4}\rm \ K$ and luminosity of $10^{43.4}\rm \ erg\ s^{-1}$, which are fitted from the multi-wavelength light curves. Solar abundances and a turbulent velocity of $1000$\,km\,s$^{-1}$ were assumed. Each calculation was terminated at a hydrogen column density of $N_{\rm H}=10^{25.5}$\,cm$^{-2}$. A constant gas density was adopted as the calculated layer is geometrically thin relative to the inner radius. Figure~\ref{fig:cloudy} qualitatively suggests that the observed optical \civ\, emission favors dense gas exposed to strong EUV illumination.


Two major limitations should be noted for the \texttt{CLOUDY} simulation of the optically thick TDE outflows. First, the calculation treats a photoionized layer illuminated by a prescribed continuum, without self-consistently following the repeated electron scattering and radiation reprocessing that shape the emergent continuum and line emission. Second, the model assumes a static envelope with prescribed turbulent broadening. More realistic configurations include an optically thick disk wind with a polar funnel \citep{Dai2018} or an outflow driven by stream--stream collisions \citep{Jiang2016,Lu2020}. The combined effect of photoionization, continuum fluorescence, and photon trapping depends on the local EUV radiation field, column density, and velocity structure. Therefore, a self-consistent treatment of radiative transfer \citep{Thomsen2026}, coupled to gas dynamics and incorporating absorption, ionization, and scattering, is needed to characterize the envelope structure and identify the regions responsible for \civ\, emission.

\subsection{Implications for TDEs and Future Prospects}
Modern radiative-transfer simulations of TDEs have reproduced key features of the observed continuum \citep{Dai2018,Thomsen2022}, the hydrogen and helium emission lines \citep{Roth2018,Parkinson2022}, with recent models also reproducing Bowen fluorescence features \citep{Thomsen2026}. However, a unified description that also accounts for carbon lines remains less developed. Among models that include carbon, the test calculations of \citet{Thomsen2026} produce \civ~$\lambda5805$ in emission when using an expanded atomic dataset, but in absorption when using a simplified dataset. This sensitivity to the adopted atomic model motivates a more complete treatment of the relevant energy levels and transitions to constrain the physical conditions in the \civ\ emission region. For the treatment of the UV \civ~1549 line, previous calculations have adopted a two-level approximation for the relevant transitions \citep{Parkinson2022,Parkinson2025}, or modeled emitting gas with a low electron-scattering optical depth \citep{Saxton2018}. The identification of \civ\ $\lambda5805$ emission motivates a joint treatment of the optical and UV \civ\ lines in the optically thick envelope \citep{Yang2017}, including the combined effect from both the continuum fluorescence and recombination cascades.

In previous TDE emission line fitting, optical \civ\, emission may have been overlooked, as TDE line widths approaching $10^{4}$\,km\,s$^{-1}$ can blend the doublet with neighboring \ion{He}{1}~$\lambda5876$, and host-galaxy Na~I~D absorption further complicates line identification. Consequently, this feature could be misinterpreted as blueshifted \ion{He}{1}~$\lambda5876$ or a candidate redshifted [\ion{N}{2}]~$\lambda5754$. In TDE~2025abcr, the narrower line profiles, with characteristic widths of $\sim3000$\,km\,s$^{-1}$, and weak host-galaxy contamination at its off-nuclear position facilitate identification of the doublet. Future spectroscopic monitoring of TDEs with resolution $R\gtrsim1000$ and higher SNR could also facilitate the identification of \civ~$\lambda5805$ emission with larger linewidth. Our preliminary visual inspection of archival TDE spectra reveals several candidate features \citep[e.g.][]{Leloudas2019}, although detailed spectral fitting is needed to assess their identification.

In conclusion, the off-nuclear TDE 2025abcr exhibits a relatively small blackbody radius, narrower broad lines, and the Bowen fluorescence lines. These properties motivate a compact, dense reprocessing-region interpretation, consistent with the proposed excitation mechanism for the \civ~$\lambda5805$ emission line. Quantitative constraints on the gas conditions will require detailed radiative-transfer modeling, high-quality multi-epoch spectra, and contemporaneous UV spectroscopy covering the \civ~$\lambda1549$ and C~III]~$\lambda1908$. A systematically observed sample of optical \civ-emitting TDEs from Legacy Survey of Space and Time (LSST) and Wide Field Survey Telescope (WFST) would establish whether these events preferentially exhibit compact, hot photospheres. Combined with independent estimates of BH and disrupted-star masses, such a sample could test how these masses influence the conditions that favor optical \civ\, emission, also providing a basis for comparing the BH and stellar populations probed by nuclear and off-nuclear TDEs. 

We thank T.G. Wang for the useful discussion. HXG acknowledges support from the National Natural Science Foundation of China (Nos. 12473018 and 12522304) and the Overseas Center Platform Projects of the Chinese Academy of Sciences (No. 178GJHZ2023184MI). WWZ is supported by the Shanghai Natural Science Foundation Youth Project (Grant No. 25ZR1402546), Strategic Priority Research Program of the Chinese Academy of Sciences (Grant No. XDB0800302), and the National Key Research and Development Program of China (Grant No. 2025YFA1614102). WWZ and HXG are supported by the National Key R\&D Program of China (Grant No. 2022YFF0503402 and 2023YFA1607903), and Future Network Partner Program, CAS (Grant No.018GJHZ2022029FN). MFG is supported by the Shanghai Pilot Program for Basic Research-Chinese Academy of Science, Shanghai Branch (Grant No. JCYJ-SHFY-2021-013), the National SKA Program of China (Grant No. 2022SKA0120102), the science research grants from the China Manned Space Project (Grant No. CMSCSST-2021-A06), and the Original Innovation Program of the Chinese Academy of Sciences 715 (Grant No. 085021002). ACG is partially supported by the CAS ``President's International Fellowship Initiative (PIFI)`` with Grant No. 2026PVA0040.

This work is based on observations collected at the European Southern Observatory under ESO programme 116.2APM.001. This work is partly based on data obtained with the instrument HiPERCAM, built by the Universities of Sheffield, Warwick and Durham, the UK Astronomy Technology Centre, and the Instituto de Astrofísica de Canarias. Development of HiPERCAM was funded by the European Research Council, and its operations and enhancements by the Science and Technology Facilities Council. This work makes use of observations from the Las Cumbres Observatory global telescope network. This paper is based on observations made with the MuSCAT instruments, developed by the Astrobiology Center (ABC) in Japan, the University of Tokyo, and Las Cumbres Observatory (LCOGT). MuSCAT3 was developed with financial support by JSPS KAKENHI (JP18H05439) and JST PRESTO (JPMJPR1775), and is located at the Faulkes Telescope North on Maui, HI (USA), operated by LCOGT. MuSCAT4 was developed with financial support provided by the Heising-Simons Foundation (grant 2022-3611), JST grant number JPMJCR1761, and the ABC in Japan, and is located at the Faulkes Telescope South at Siding Spring Observatory (Australia), operated by LCOGT. This paper is based on observations made with the Nordic Optical Telescope, owned in collaboration by the University of Turku and Aarhus University, and operated jointly by Aarhus University, the University of Turku and the University of Oslo, representing Denmark, Finland and Norway, the University of Iceland and Stockholm University at the Observatorio del Roque de los Muchachos, La Palma, Spain, of the Instituto de Astrofisica de Canarias. The data presented here were obtained in part with ALFOSC, which is provided by the Instituto de Astrofisica de Andalucia (IAA) under a joint agreement with the University of Copenhagen and NOT. This work made use of data supplied by the UK Swift Science Data Centre at the University of Leicester. This work is based on observations obtained with the Samuel Oschin Telescope 48-inch and the 60-inch Telescope at the Palomar Observatory as part of the Zwicky Transient Facility project. ZTF is supported by the National Science Foundation under Grants No. AST-1440341 and AST-2034437 and a collaboration including current partners Caltech, IPAC, the Oskar Klein Center at Stockholm University, the University of Maryland, University of California, Berkeley, the University of Wisconsin at Milwaukee, University of Warwick, Ruhr University, Cornell University, Northwestern University and Drexel University. Operations are conducted by COO, IPAC, and UW. This work makes use of data from the ZTF Forced Photometry Service. The ZTF forced-photometry service was funded under the Heising-Simons Foundation grant \#12540303 (PI: Graham).

\facilities{ZTF, VLT (FORS2), GTC (HiPERCAM), LCOGT (MuSCAT3, MuSCAT4), Swift (UVOT, XRT), NOT (ALFOSC)}

\software{
SExtractor \citep{Bertin1996},
PSFEx \citep{Bertin2011},
SCAMP \citep{Bertin2006},
SWarp \citep{Bertin2002},
HOTPANTS \citep{Becker2015},
CLOUDY \citep{Chatzikos2023}
          }

\bibliography{main}{}
\bibliographystyle{aasjournalv7.1}

\end{document}